\documentstyle[epic,eepic]{article}
\title{The complexity of planar graph choosability
\thanks{This paper forms part of a Ph.D. thesis written by the author
under the supervision of Prof. N. Alon and Prof. M. Tarsi
in Tel Aviv University.}}
\author{Shai Gutner \\
	Department of Computer Science \\
	School of Mathematical Sciences \\
	Raymond and Beverly Sackler Faculty of Exact Sciences \\
	Tel Aviv University, Tel Aviv, Israel}
\date{}
\newtheorem{theo}{Theorem}[section]

\newtheorem{lemma}[theo]{Lemma}

\newtheorem{defi}[theo]{Definition}

\begin{document}
\maketitle
\setcounter{page}{0}
\begin{abstract}
A graph $G$ is {\em $k$-choosable} if for every assignment of a 
set~$S(v)$ of $k$~colors to every vertex~$v$ of~$G$, there is a proper 
coloring of $G$ that assigns to each vertex~$v$ a color from~$S(v)$.
We consider the complexity of deciding whether a given graph is $k$-choosable
for some constant~$k$. 
In particular, it is shown that deciding whether a given planar graph is 
$4$-choosable is NP-hard, and so is the problem of deciding whether 
a given planar triangle-free graph is $3$-choosable.
We also obtain simple constructions of a planar graph which is not 4-choosable
and a planar triangle-free graph which is not 3-choosable.
\end{abstract}
\newpage
\section{Introduction}

All graphs considered here are finite, undirected and simple         
(i.e., have no loops and no parallel edges).
If~$G=(V,E)$ is a graph, and $f$ is a function that assigns to each 
vertex~$v$ of~$G$ a positive integer $f(v)$, we say that $G$ is $f$-choosable
if for every assignment of sets of integers $S(v) \subseteq Z$ for all       
vertices $v \in V$, where $|S(v)| = f(v)$ for all $v$, there is a proper
vertex coloring $c:V \mapsto Z$ so that $c(v) \in S(v)$ for all $v \in V$.
The graph $G$ is {\em $k$-choosable} if it is $f$-choosable for the        
constant function $f(v) \equiv k$. The {\em choice number} of $G$, 
denoted $ch(G)$, is the minimum integer $k$ so that $G$ is $k$-choosable.

The study of choice numbers of graphs was initiated by Vizing in~\cite{V}
and by Erd\H{o}s, Rubin and Taylor in~\cite{ERT}. 
A characterization of all $2$-choosable graphs is given in~\cite{ERT}.
If $G$ is a connected graph, the {\em core} of~$G$ is the graph obtained
from $G$ by repeatedly deleting vertices of degree $1$ until there is no
such vertex.
\begin{theo}[\cite{ERT}]
\label{t11}
A simple graph is $2$-choosable if and only if the core of each connected
component of it is either a single vertex, or an even cycle, or a graph
consisting of two vertices with three even internally disjoint paths 
between them, where the length of at least two of the paths is 
exactly $2$.
\end{theo}

In the present paper we consider the complexity of deciding whether  
a given graph is $k$-choosable for some constant $k$. It is shown  
in~\cite{ERT} that the following problem is $\Pi_2^p$-complete:
(for terminology see~\cite{GJ})

\vspace{3mm}
\noindent
{\bf BIPARTITE GRAPH (2,3)-CHOOSABILITY (BG (2,3)-CH)} \\
INSTANCE: A bipartite graph $G=(V,E)$ and 
a function $f:V \mapsto \{2,3\}$. \\
QUESTION: Is $G$ $f$-choosable?

\vspace{3mm}
\noindent
Consider the following decision problem:

\vspace{3mm}
\noindent
{\bf BIPARTITE GRAPH $k$-CHOOSABILITY (BG $k$-CH)} \\
INSTANCE: A bipartite graph $G$. \\ 
QUESTION: Is $G$ $k$-choosable?

\vspace{3mm}
\noindent
It is proved in~\cite{GT} that this problem is $\Pi_2^p$-complete for
every constant $k \geq 3$. 
It follows easily from Theorem~\ref{t11} that the case $k=2$
is solvable in  polynomial time. 

The following results are known concerning the choice numbers of
planar graphs:
\begin{theo}[\cite{T1}]
\label{t12}
Every planar graph is $5$-choosable.
\end{theo}
\begin{theo}[\cite{Vo}]
\label{t13}
There exists a planar graph (with $238$ vertices) which is not
$4$-choosable.
\end{theo}
\begin{theo}[\cite{AT}]
\label{t14}
Every bipartite planar graph is $3$-choosable.
\end{theo}
\begin{theo}[\cite{Vo2}]
\label{t15}
There exists a planar triangle-free graph (with $166$ vertices) which
is not 3-choosable.
\end{theo}
\begin{theo}[\cite{T2}]
\label{t16}
Every planar graph with girth $5$ is $3$-choosable.
\end{theo}
The following two theorems improve upon Theorems~\ref{t13} and~\ref{t15}  
and use much simpler constructions.
\begin{theo}
\label{t17}
There exists a planar graph with $75$ vertices which is not 
$4$-choosable.
\end{theo}
\begin{theo}
\label{t18}
There exists a planar triangle-free graph with $164$ vertices which is   
not $3$-choosable.
\end{theo}

It follows easily from Theorems~\ref{t11} and~\ref{t14} that the   
choice number of a given bipartite planar graph can be determined in
polynomial time. 
Consider the following decision problems:

\vspace{3mm}
\noindent
{\bf BIPARTITE PLANAR GRAPH $(2,3)$-CHOOSABILITY (BPG $(2,3)$-CH)} \\
INSTANCE: A bipartite planar graph $G=(V,E)$ and
a function $f:V \mapsto \{2,3\}$. \\
QUESTION: Is $G$ $f$-choosable?

\vspace{3mm}
\noindent
{\bf PLANAR TRIANGLE-FREE GRAPH $3$-CHOOSABILITY (PTFG $3$-CH)} \\
INSTANCE: A planar triangle-free graph $G$. \\
QUESTION: Is $G$ $3$-choosable?

\vspace{3mm}
\noindent
{\bf PLANAR GRAPH $4$-CHOOSABILITY (PG $4$-CH)} \\
INSTANCE: A planar graph $G$. \\
QUESTION: Is $G$ $4$-choosable?

\vspace{3mm}
\noindent
{\bf UNION OF TWO FORESTS $3$-CHOOSABILITY (U2F $3$-CH)} \\
INSTANCE: Two forests $F_1$ and $F_2$ with $V(F_1)=V(F_2)$. \\
QUESTION: Is the union of $F_1$ and $F_2$ $3$-choosable?

\vspace{3mm}
\noindent
We prove the following results:
\begin{theo}
\label{t19}
{\bf BIPARTITE PLANAR GRAPH (2,3)-CHOOSABILITY}
is $\Pi_2^p$-complete.
\end{theo}
\begin{theo}
\label{t110}
{\bf PLANAR TRIANGLE-FREE GRAPH $3$-CHOOSABILITY} is
$\Pi_2^p$-complete.
\end{theo}
\begin{theo}
\label{t111}
{\bf PLANAR GRAPH $4$-CHOOSABILITY} is $\Pi_2^p$-complete.
\end{theo}

The decision problem {\bf U2F $3$-CH} was formulated by 
M. Stiebitz~\cite{St} in light of the fact that every
planar triangle-free graph is the union of two forests. 
The following Theorem can be derived easily from the constructions
used in the proofs of Theorems~\ref{t19} and~\ref{t110}.
\begin{theo}
\label{t112}
{\bf UNION OF TWO FORESTS $3$-CHOOSABILITY} is $\Pi_2^p$-complete.
\end{theo} 
 
The rest of the paper is organized as follows. 
In section 2 we prove
Theorems~\ref{t17} and~\ref{t18}. 
The $\Pi_2^p$-completeness proof
of the decision problem {\bf BG (2,3)-CH} taken from~\cite{ERT}
forms the basis for the
proof of Theorem~\ref{t19} given in section 3.
Section 4 contains the proofs of Theorems~\ref{t110} and~\ref{t111}.
 
\section{Two planar graphs}

In this section we construct two planar graphs in order to
prove Theorems~\ref{t17} and~\ref{t18}.

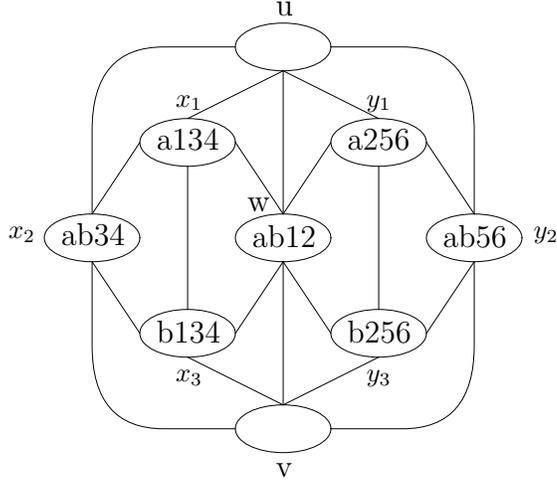
\begin{figure}[t]
%\centerline{\input{fig1.eepic}}
$$
\setlength{\unitlength}{0.0125in}
\begin{picture}(250,214)(0,-10)
\put(35,100){\ellipse{40}{20}}
\put(115,100){\ellipse{40}{20}}
\put(155,140){\ellipse{40}{20}}
\put(195,100){\ellipse{40}{20}}
\put(155,60){\ellipse{40}{20}}
\put(115,180){\ellipse{40}{20}}
\put(115,20){\ellipse{40}{20}}
\put(75,140){\ellipse{40}{20}}
\put(75,60){\ellipse{40}{20}}
\path(155,130)(155,70)
\path(195,90)(175,60)
\path(175,140)(195,110)
\path(135,140)(115,110)
\path(115,90)(135,60)
\path(95,140)(115,110)
\path(55,140)(35,110)
\path(35,90)(55,60)
\path(115,90)(95,60)
\path(75,130)(75,70)
\path(115,170)(115,110)
\path(115,90)(115,30)
\path(115,170)(155,150)
\path(115,170)(75,150)
\path(75,50)(115,30)
\path(155,50)(115,30)
\spline(95,180)
(35,180)(35,110)
\spline(135,180)
(195,180)(195,110)
\spline(35,90)
(35,20)(95,20)
\spline(195,90)
(195,20)(135,20)
\put(22,97){\makebox(0,0)[lb]{\raisebox{0pt}[0pt][0pt]{\shortstack[l]{{\twlrm ab34}}}}}
\put(62,137){\makebox(0,0)[lb]{\raisebox{0pt}[0pt][0pt]{\shortstack[l]{{\twlrm a134}}}}}
\put(142,137){\makebox(0,0)[lb]{\raisebox{0pt}[0pt][0pt]{\shortstack[l]{{\twlrm a256}}}}}
\put(102,96){\makebox(0,0)[lb]{\raisebox{0pt}[0pt][0pt]{\shortstack[l]{{\twlrm ab12}}}}}
\put(182,96){\makebox(0,0)[lb]{\raisebox{0pt}[0pt][0pt]{\shortstack[l]{{\twlrm ab56}}}}}
\put(62,56){\makebox(0,0)[lb]{\raisebox{0pt}[0pt][0pt]{\shortstack[l]{{\twlrm b134}}}}}
\put(142,56){\makebox(0,0)[lb]{\raisebox{0pt}[0pt][0pt]{\shortstack[l]{{\twlrm b256}}}}}
\put(112,0){\makebox(0,0)[lb]{\raisebox{0pt}[0pt][0pt]{\shortstack[l]{{\twlrm v}}}}}
\put(112,193){\makebox(0,0)[lb]{\raisebox{0pt}[0pt][0pt]{\shortstack[l]{{\twlrm u}}}}}
\put(150,155){\makebox(0,0)[lb]{\raisebox{0pt}[0pt][0pt]{\shortstack[l]{{\twlrm $y_1$}}}}}
\put(150,40){\makebox(0,0)[lb]{\raisebox{0pt}[0pt][0pt]{\shortstack[l]{{\twlrm $y_3$}}}}}
\put(0,100){\makebox(0,0)[lb]{\raisebox{0pt}[0pt][0pt]{\shortstack[l]{{\twlrm $x_2$}}}}}
\put(220,100){\makebox(0,0)[lb]{\raisebox{0pt}[0pt][0pt]{\shortstack[l]{{\twlrm $y_2$}}}}}
\put(100,111){\makebox(0,0)[lb]{\raisebox{0pt}[0pt][0pt]{\shortstack[l]{{\twlrm w}}}}}
\put(70,155){\makebox(0,0)[lb]{\raisebox{0pt}[0pt][0pt]{\shortstack[l]{{\twlrm $x_1$}}}}}
\put(70,40){\makebox(0,0)[lb]{\raisebox{0pt}[0pt][0pt]{\shortstack[l]{{\twlrm $x_3$}}}}}
\end{picture}
$$
\vspace*{-15pt}
\caption[]{\label{f1}The graph $W_1$.}
\end{figure}

\noindent
{\bf Proof of Theorem~\ref{t17}}\,
The graph $H$ is constructed as follows:
We take the disjoint union of the 
graphs~$\{G_i:1 \leq i \leq 12\}$, where each $G_i$ is a copy of the 
graph $W_1$ in Fig.~\ref{f1}.
All the $12$ vertices named~$u$ are identified, as
well as all the $12$ vertices named~$v$. 
The edge $(u,v)$ is added
to obtain the graph $H$, which is obviously planar.
We claim that the graph $H$ is not $4$-choosable. 
To prove this, take
$S(u) = S(v) = \{7,8,9,10\}$. 
Denote $A = \{(a,b) \in S(u) \times S(v)| a \neq b \}$, 
then surely $|A| = 12$. 
With every $i$, $1 \leq i \leq 12$ we associate a different element 
$p_i = (a,b) \in A$, and define the sets of every vertex of $G_i$
except for $u$ and $v$ to be as in Fig.~\ref{f1}. 
It can be easily verified that there is no proper vertex coloring
for this assignment, and therefore $H$ is not $4$-choosable.
To see this, suppose the vertex $u$ is colored with the color $a$
and the vertex $v$ is colored with the color $b$, 
where $(a,b)=p_i \in A$.
The vertex $w$ in the graph $G_i$ can be colored with either the
color $1$ or the color $2$, and in both cases the coloring
in the graph $G_i$ cannot be completed.

We now construct a planar graph $H'$ which is not $4$-choosable 
and has fewer
vertices than $H$. The graph~$H'$ is obtained from $H$ by identifying
the vertex $y_2$ of $G_i$ with the vertex $x_2$ of $G_{i+1}$ 
for every~$i$, $1 \leq i < 12$. We claim that $H'$ is not $4$-choosable.
The previous definitions of $S(u)$, $S(v)$, $A$ and $p_i$ are used.  
For every~$i$, $1 \leq i < 12$ we do the following:
Denote $p_i = (a,b)$ and $p_{i+1} = (c,d)$. 
The set of the vertex~$y_2$ of~$G_i$ (which is the same as the set of
the vertex $x_2$ of $G_{i+1}$) is chosen so that it contains the colors
$a$,$b$,$c$ and~$d$ 
(and maybe other colors if $p_i$ and $p_{i+1}$ are not disjoint).
In the same manner as before, we conclude that $H'$ is not $4$-choosable.
The graph $H'$ is planar and has~$2+12*7-11=75$ vertices. $\Box$

\begin{figure}[t]
%\centerline{\input{fig2.eepic}}
$$
\setlength{\unitlength}{0.0125in}
\begin{picture}(512,338)(0,-10)
\put(196,161){\ellipse{40}{20}}
\put(176,201){\ellipse{40}{20}}
\put(116,201){\ellipse{40}{20}}
\put(116,121){\ellipse{40}{20}}
\put(176,241){\ellipse{40}{20}}
\put(176,81){\ellipse{40}{20}}
\put(76,241){\ellipse{40}{20}}
\put(76,81){\ellipse{40}{20}}
\put(316,161){\ellipse{40}{20}}
\put(336,201){\ellipse{40}{20}}
\put(336,121){\ellipse{40}{20}}
\put(396,201){\ellipse{40}{20}}
\put(396,121){\ellipse{40}{20}}
\put(336,241){\ellipse{40}{20}}
\put(336,81){\ellipse{40}{20}}
\put(436,241){\ellipse{40}{20}}
\put(176,121){\ellipse{40}{20}}
\put(436,81){\ellipse{40}{20}}
\put(256,161){\ellipse{40}{20}}
\put(256,301){\ellipse{40}{20}}
\put(256,21){\ellipse{40}{20}}
\path(176,191)(196,171)
\path(196,151)(176,131)
\path(156,201)(136,201)
\path(156,121)(136,121)
\path(116,191)(116,131)
\path(196,241)(256,171)
\path(96,241)(156,241)
\path(176,231)(176,211)
\path(236,161)(216,161)
\path(176,111)(176,91)
\path(76,231)(76,91)
\path(96,81)(156,81)
\path(276,161)(296,161)
\path(336,191)(316,171)
\path(316,151)(336,131)
\path(356,201)(376,201)
\path(356,121)(376,121)
\path(396,191)(396,131)
\path(316,241)(256,171)
\path(256,151)(316,81)
\path(336,231)(336,211)
\path(336,111)(336,91)
\path(356,241)(416,241)
\path(436,231)(436,91)
\path(416,81)(356,81)
\path(76,251)(236,301)
\path(256,291)(256,171)
\path(436,251)(276,301)
\path(336,71)(256,31)
\path(107,210)(85,232)
\path(107,112)(85,90)
\path(427,232)(405,210)
\path(427,90)(405,112)
\path(176,71)(256,31)
\path(256,151)(196,81)
\path(56,241)	(53.498,237.329)
	(51.061,233.713)
	(48.688,230.153)
	(46.379,226.647)
	(44.133,223.195)
	(41.951,219.795)
	(39.832,216.446)
	(37.776,213.148)
	(35.782,209.900)
	(33.851,206.701)
	(31.982,203.549)
	(30.175,200.445)
	(28.430,197.386)
	(26.747,194.373)
	(25.125,191.403)
	(23.564,188.477)
	(22.064,185.594)
	(20.624,182.752)
	(19.245,179.950)
	(17.926,177.188)
	(16.667,174.465)
	(15.468,171.780)
	(14.328,169.131)
	(13.248,166.519)
	(12.226,163.941)
	(11.264,161.398)
	(9.514,156.410)
	(7.997,151.547)
	(6.710,146.803)
	(5.652,142.170)
	(4.822,137.640)
	(4.216,133.207)
	(3.834,128.862)
	(3.673,124.600)
	(3.732,120.411)
	(4.009,116.289)
	(4.501,112.227)
	(5.208,108.218)
	(6.128,104.253)
	(7.257,100.326)
	(8.596,96.430)
	(10.141,92.556)
	(11.892,88.698)
	(13.845,84.848)
	(16.000,81.000)

\path(16,81)	(17.575,78.369)
	(19.183,75.798)
	(22.504,70.838)
	(25.972,66.118)
	(29.598,61.634)
	(33.392,57.384)
	(37.363,53.365)
	(41.522,49.575)
	(45.879,46.010)
	(50.444,42.668)
	(55.227,39.546)
	(57.703,38.067)
	(60.238,36.642)
	(62.832,35.270)
	(65.487,33.952)
	(68.204,32.687)
	(70.985,31.474)
	(73.830,30.314)
	(76.741,29.205)
	(79.719,28.148)
	(82.765,27.143)
	(85.881,26.188)
	(89.068,25.284)
	(92.327,24.430)
	(95.659,23.626)
	(99.066,22.872)
	(102.549,22.167)
	(106.109,21.510)
	(109.747,20.902)
	(113.466,20.343)
	(117.265,19.831)
	(121.146,19.366)
	(125.111,18.949)
	(129.160,18.579)
	(133.296,18.255)
	(137.518,17.977)
	(141.830,17.745)
	(146.231,17.558)
	(150.723,17.416)
	(155.307,17.319)
	(159.985,17.267)
	(164.757,17.258)
	(169.626,17.293)
	(174.592,17.372)
	(177.112,17.427)
	(179.656,17.494)
	(182.226,17.570)
	(184.821,17.658)
	(187.441,17.756)
	(190.086,17.864)
	(192.757,17.984)
	(195.454,18.113)
	(198.177,18.253)
	(200.925,18.403)
	(203.700,18.564)
	(206.502,18.735)
	(209.330,18.916)
	(212.184,19.107)
	(215.065,19.309)
	(217.974,19.520)
	(220.909,19.742)
	(223.872,19.974)
	(226.863,20.215)
	(229.881,20.467)
	(232.926,20.729)
	(236.000,21.000)

\path(456,241)	(458.501,237.329)
	(460.939,233.713)
	(463.312,230.153)
	(465.621,226.647)
	(467.867,223.195)
	(470.049,219.795)
	(472.168,216.446)
	(474.224,213.148)
	(476.218,209.900)
	(478.149,206.701)
	(480.017,203.549)
	(481.824,200.445)
	(483.569,197.386)
	(485.252,194.373)
	(486.875,191.403)
	(488.436,188.477)
	(489.936,185.594)
	(491.375,182.752)
	(492.754,179.950)
	(494.073,177.188)
	(495.332,174.465)
	(496.531,171.780)
	(497.671,169.131)
	(498.751,166.519)
	(499.773,163.941)
	(500.735,161.398)
	(502.485,156.410)
	(504.002,151.547)
	(505.289,146.803)
	(506.347,142.170)
	(507.178,137.640)
	(507.783,133.207)
	(508.165,128.862)
	(508.326,124.600)
	(508.267,120.411)
	(507.991,116.289)
	(507.498,112.227)
	(506.791,108.218)
	(505.872,104.253)
	(504.742,100.326)
	(503.404,96.430)
	(501.859,92.556)
	(500.108,88.698)
	(498.155,84.848)
	(496.000,81.000)

\path(496,81)	(494.425,78.369)
	(492.817,75.798)
	(489.496,70.838)
	(486.028,66.118)
	(482.403,61.634)
	(478.609,57.384)
	(474.638,53.366)
	(470.479,49.575)
	(466.122,46.010)
	(461.557,42.669)
	(456.774,39.547)
	(454.298,38.067)
	(451.763,36.642)
	(449.169,35.271)
	(446.514,33.953)
	(443.797,32.687)
	(441.017,31.475)
	(438.172,30.314)
	(435.261,29.206)
	(432.283,28.149)
	(429.236,27.143)
	(426.121,26.189)
	(422.934,25.285)
	(419.675,24.431)
	(416.342,23.627)
	(412.936,22.872)
	(409.453,22.167)
	(405.893,21.511)
	(402.254,20.903)
	(398.536,20.343)
	(394.737,19.831)
	(390.855,19.367)
	(386.891,18.950)
	(382.841,18.579)
	(378.706,18.255)
	(374.483,17.977)
	(370.171,17.745)
	(365.770,17.558)
	(361.278,17.417)
	(356.694,17.320)
	(352.016,17.267)
	(347.244,17.259)
	(342.375,17.294)
	(337.409,17.372)
	(334.889,17.428)
	(332.344,17.494)
	(329.775,17.571)
	(327.180,17.658)
	(324.560,17.756)
	(321.914,17.865)
	(319.243,17.984)
	(316.547,18.113)
	(313.824,18.253)
	(311.075,18.404)
	(308.300,18.564)
	(305.499,18.735)
	(302.671,18.916)
	(299.816,19.108)
	(296.935,19.309)
	(294.026,19.521)
	(291.091,19.742)
	(288.128,19.974)
	(285.138,20.216)
	(282.120,20.467)
	(279.074,20.729)
	(276.000,21.000)

\put(167,237){\makebox(0,0)[lb]{\raisebox{0pt}[0pt][0pt]{\shortstack[l]{{\twlrm 135}}}}}
\put(107,197){\makebox(0,0)[lb]{\raisebox{0pt}[0pt][0pt]{\shortstack[l]{{\twlrm 389}}}}}
\put(168,197){\makebox(0,0)[lb]{\raisebox{0pt}[0pt][0pt]{\shortstack[l]{{\twlrm 589}}}}}
\put(108,117){\makebox(0,0)[lb]{\raisebox{0pt}[0pt][0pt]{\shortstack[l]{{\twlrm 789}}}}}
\put(187,157){\makebox(0,0)[lb]{\raisebox{0pt}[0pt][0pt]{\shortstack[l]{{\twlrm 189}}}}}
\put(67,78){\makebox(0,0)[lb]{\raisebox{0pt}[0pt][0pt]{\shortstack[l]{{\twlrm 367}}}}}
\put(166,77){\makebox(0,0)[lb]{\raisebox{0pt}[0pt][0pt]{\shortstack[l]{{\twlrm b16}}}}}
\put(167,117){\makebox(0,0)[lb]{\raisebox{0pt}[0pt][0pt]{\shortstack[l]{{\twlrm 689}}}}}
\put(307,157){\makebox(0,0)[lb]{\raisebox{0pt}[0pt][0pt]{\shortstack[l]{{\twlrm 289}}}}}
\put(327,197){\makebox(0,0)[lb]{\raisebox{0pt}[0pt][0pt]{\shortstack[l]{{\twlrm 589}}}}}
\put(326,237){\makebox(0,0)[lb]{\raisebox{0pt}[0pt][0pt]{\shortstack[l]{{\twlrm 245}}}}}
\put(386,197){\makebox(0,0)[lb]{\raisebox{0pt}[0pt][0pt]{\shortstack[l]{{\twlrm 489}}}}}
\put(426,77){\makebox(0,0)[lb]{\raisebox{0pt}[0pt][0pt]{\shortstack[l]{{\twlrm 467}}}}}
\put(327,77){\makebox(0,0)[lb]{\raisebox{0pt}[0pt][0pt]{\shortstack[l]{{\twlrm b26}}}}}
\put(327,117){\makebox(0,0)[lb]{\raisebox{0pt}[0pt][0pt]{\shortstack[l]{{\twlrm 689}}}}}
\put(387,117){\makebox(0,0)[lb]{\raisebox{0pt}[0pt][0pt]{\shortstack[l]{{\twlrm 789}}}}}
\put(247,158){\makebox(0,0)[lb]{\raisebox{0pt}[0pt][0pt]{\shortstack[l]{{\twlrm a12}}}}}
\put(66,238){\makebox(0,0)[lb]{\raisebox{0pt}[0pt][0pt]{\shortstack[l]{{\twlrm ab3}}}}}
\put(426,238){\makebox(0,0)[lb]{\raisebox{0pt}[0pt][0pt]{\shortstack[l]{{\twlrm ab4}}}}}
\put(253,317){\makebox(0,0)[lb]{\raisebox{0pt}[0pt][0pt]{\shortstack[l]{{\twlrm u}}}}}
\put(253,0){\makebox(0,0)[lb]{\raisebox{0pt}[0pt][0pt]{\shortstack[l]{{\twlrm v}}}}}
\put(171,256){\makebox(0,0)[lb]{\raisebox{0pt}[0pt][0pt]{\shortstack[l]{{\twlrm $x_1$}}}}}
\put(71,256){\makebox(0,0)[lb]{\raisebox{0pt}[0pt][0pt]{\shortstack[l]{{\twlrm $x_2$}}}}}
\put(252,137){\makebox(0,0)[lb]{\raisebox{0pt}[0pt][0pt]{\shortstack[l]{{\twlrm w}}}}}
\put(71,61){\makebox(0,0)[lb]{\raisebox{0pt}[0pt][0pt]{\shortstack[l]{{\twlrm $x_3$}}}}}
\put(171,61){\makebox(0,0)[lb]{\raisebox{0pt}[0pt][0pt]{\shortstack[l]{{\twlrm $x_4$}}}}}
\put(331,256){\makebox(0,0)[lb]{\raisebox{0pt}[0pt][0pt]{\shortstack[l]{{\twlrm $y_1$}}}}}
\put(431,256){\makebox(0,0)[lb]{\raisebox{0pt}[0pt][0pt]{\shortstack[l]{{\twlrm $y_2$}}}}}
\put(431,61){\makebox(0,0)[lb]{\raisebox{0pt}[0pt][0pt]{\shortstack[l]{{\twlrm $y_3$}}}}}
\put(331,61){\makebox(0,0)[lb]{\raisebox{0pt}[0pt][0pt]{\shortstack[l]{{\twlrm $y_4$}}}}}
\end{picture}
$$
\vspace*{-15pt}
\caption[]{\label{f2}The graph $W_2$.}
\end{figure}
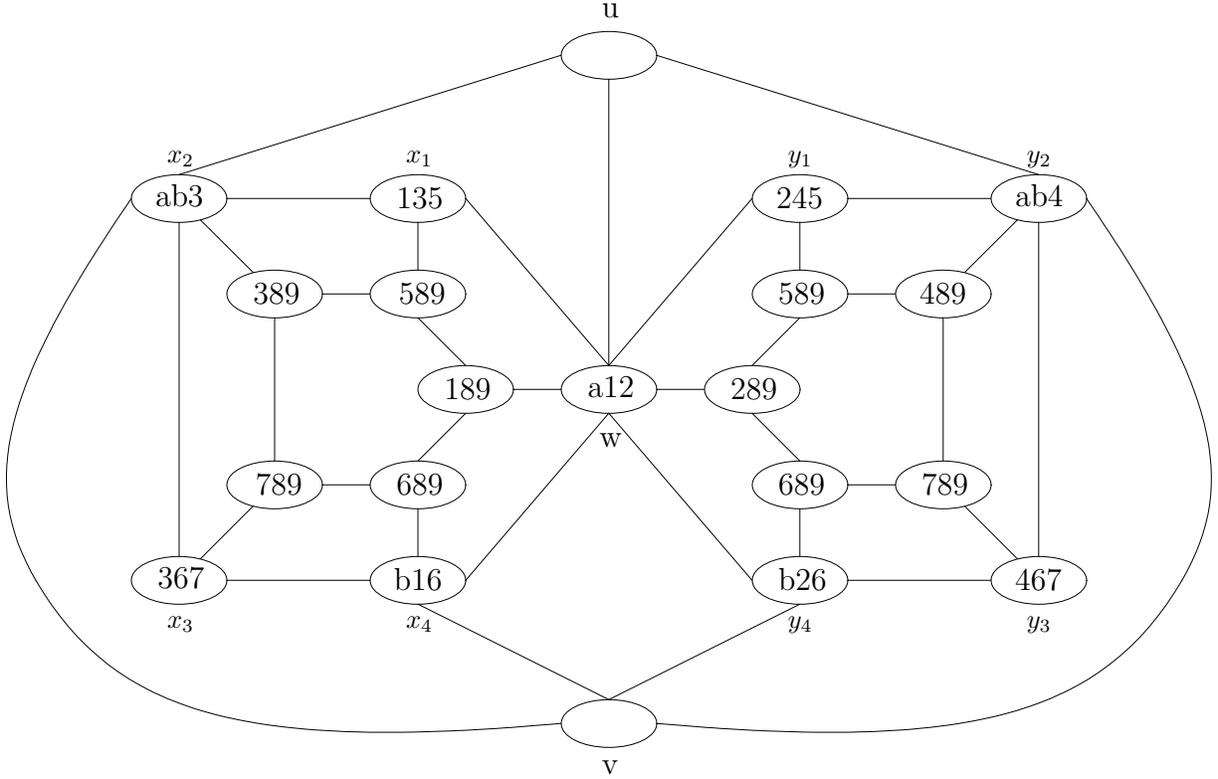

\noindent
{\bf Proof of Theorem~\ref{t18}}\,
The graph $H$ is constructed as follows:
We take the disjoint union of the 
graphs~$\{G_i:1 \leq i \leq 9\}$, where each $G_i$ is a copy of the 
graph $W_2$ in Fig.~\ref{f2}. 
All the $9$ vertices named~$u$ are identified, as
well as all the $9$ vertices named~$v$, to obtain the planar
triangle-free graph $H$. 
We claim that the graph $H$ is not $3$-choosable. 
To prove this, take
$S(u) = \{10,11,12\}$ and $S(v) = \{13,14,15\}$. 
With every $i$, $1 \leq i \leq 9$ we associate a different element 
$(a,b) \in S(u) \times S(v)$, and define the sets of every vertex of $G_i$
except for~$u$ and~$v$ to be as in Fig.~\ref{f2}. 
As in the proof of Theorem~\ref{t17},
we conclude that~$H$ is not $3$-choosable.

We now construct a planar triangle-free graph $H'$ which is not $3$-choosable 
and has fewer
vertices than~$H$. The graph $H'$ is obtained from $H$ by identifying
the vertex $y_2$ of $G_i$ with the vertex $x_2$ of $G_{i+1}$ 
for every~$i$, $1 \leq i \leq 9$
(indices taken modulo 9). We claim that $H'$ is not $3$-choosable.
The previous definitions of~$S(u)$ and~$S(v)$ are used.  
Consider the following ordering of the elements of~$S(u) \times S(v)$:
$$
\{p_i\}_{i=1}^9 = (10,13),(10,14),(10,15),(11,15),(11,13),(11,14),
(12,14),(12,15),(12,13).
$$
For every~$i$, $1 \leq i \leq 9$ we do the following:
Denote $p_i = (a,b)$ and $p_{i+1} = (c,d)$. 
The set of the vertex~$y_2$ of~$G_i$ (which is the same as the set of
the vertex $x_2$ of $G_{i+1}$) is defined 
as~$\{a,b,c,d\}$ (this is a set of size $3$). 
In the same manner as before, we conclude that $H'$ is not $3$-choosable.
$H'$ is a planar triangle-free graph
and has~$2+9*18=164$~vertices. $\Box$

\section{The choosability of bipartite planar graphs}

The $\Pi_2^p$-completeness proof
of the decision problem {\bf BG (2,3)-CH} taken from~\cite{ERT}
forms the basis for the
proof of Theorem~\ref{t19} given in this section.
The ordinary Planar Satisfiability problem is well known to be 
NP-complete~(\cite{GJ},\cite{Li}).
We use a reduction from the following problem:

\vspace{3mm}
\noindent
{\bf RESTRICTED PLANAR SATISFIABILITY (RPS)} \\
INSTANCE: An expression of the form 
$(\forall U_1) \cdots (\forall U_k)(\exists V_1) \cdots (\exists V_r) \Phi$
such that
(1)~$\Phi$ is a formula in conjunctive normal form with a set $C$ of clauses
over the set $X = \{U_1, \ldots ,U_k,V_1, \ldots ,V_r\}$ of variables,
(2)~each clause involves exactly three distinct variables,
(3)~every variable occurs in at most three clauses, and
(4)~the graph 
$G_{\Phi}=(X \cup C,\{xc|x \in c \in C$ or $\overline{x} \in c \in C\})$
is planar. \\
QUESTION: Is this expression true?

\vspace{3mm}
\noindent
A similar problem is used in~\cite{KT} for proving results concerning the 
complexity of list colorings.
The same transformation used in~\cite{Li} for proving that the decision
problem Planar Quantified Boolean Formula is P-space-complete
can be used for proving that the following problem is $\Pi_2^p$-complete:

\vspace{3mm}
\noindent
{\bf ORDINARY PLANAR SATISFIABILITY (OPS)} \\
INSTANCE: An expression of the form 
$(\forall U_1) \cdots (\forall U_k)(\exists V_1) \cdots (\exists V_r) \Phi$
such that
(1)~$\Phi$ is a formula in conjunctive normal form with a set $C$ of clauses
over the set $X = \{U_1, \ldots ,U_k,V_1, \ldots ,V_r\}$ of variables,
(2)~each clause involves at most three distinct variables,
(3)~the graph 
$G_{\Phi}=(X \cup C,\{xc|x \in c \in C$ or $\overline{x} \in c \in C\})$
is planar. \\
QUESTION: Is this expression true?

\vspace{3mm}
\noindent
We apply ideas from~\cite{MP} for proving the following lemma:
\begin{lemma}
\label{l21}
{\bf RESTRICTED PLANAR SATISFIABILITY} is $\Pi_2^p$-complete.
\end{lemma}
{\bf Proof}\,
It is easy to see that {\bf RPS}$\in \Pi_2^p$.
We transform {\bf OPS} to {\bf RPS}. 
Let the expression $B$ be an instance of {\bf OPS},
and suppose that $B$ has the form
$(\forall U_1) \cdots (\forall U_k)(\exists U_{k+1}) \cdots (\exists U_{k+r}) \Phi$.
Take a planar embedding of~$G_{\Phi}$.
For every variable $V$ we do the following:
Let $(V,C_1), \ldots ,(V,C_n)$ 
be the edges adjacent to the variable~$V$ in the graph $G_{\Phi}$
in a clockwise order according to the planar embedding. 
Now introduce new variables~$V_1, \ldots ,V_n$ and clauses 
$V_i \vee \overline{V}_{i+1}$, $i=1,\ldots,n$ (indices taken modulo $n$),
and replace the literals~$V,\overline{V}$ in clauses $C_i$ by 
the literals~$V_i,\overline{V}_i$, respectively, for $i=1,\ldots,n$.
The quantified variable $V$ is replaced with the variable $V_1$ quantified
with the same quantifier.
A new quantifier block existentially quantifying the variables
$V_2, \ldots ,V_n$ is appended to the list of quantifiers.

To every clause which involves exactly two variables we 
add a new variable $V$ and insert the quantified variable $(\forall V)$ in
the beginning of the expression. In a similar manner we handle 
clauses with only one variable.
It is easily seen that the modified formula has the desired properties
and that it is true if and only if $B$ is true. $\Box$
 
\noindent
{\bf Proof of Theorem~\ref{t19}}\,
It is easy to see that {\bf BPG $(2,3)$-CH}$\in \Pi_2^p$.
We transform {\bf RPS} to {\bf BPG $(2,3)$-CH}.
Let the expression
$(\forall U_1) \cdots (\forall U_k)(\exists U_{k+1}) \cdots (\exists U_{k+r}) \Phi$,
denoted as $B$,
be an instance of {\bf RPS}. 
We shall construct a bipartite planar graph $G=(V,E)$ and 
a function $f:V \mapsto \{2,3\}$ such that $G$ is $f$-choosable
if and only if $B$ is true.
Suppose that $\Phi$ has the following form: 
$C_1 \wedge C_2 \wedge \cdots \wedge C_m$         
where each $C_i$ is of the form $(X_{i1} \vee X_{i2} \vee X_{i3})$
and each $X_{ij}$ is $U_s$ or $\overline{U}_s$.

\begin{figure}[t]
%\centerline{\input{fig3.eepic}}
$$
\setlength{\unitlength}{0.0125in}
\begin{picture}(202,115)(0,-10)
\put(70,50){\ellipse{20}{20}}
\put(130,50){\ellipse{20}{20}}
\put(10,50){\ellipse{20}{20}}
\put(70,90){\ellipse{20}{20}}
\put(70,10){\ellipse{20}{20}}
\put(130,10){\ellipse{20}{20}}
\put(190,50){\ellipse{20}{20}}
\path(60,50)(20,50)
\path(80,10)(120,10)
\path(140,50)(180,50)
\path(60,90)(20,50)
\path(20,50)(60,10)
\path(80,90)(120,50)
\path(120,50)(80,50)
\path(120,50)(80,10)
\path(180,50)(140,10)
\put(178,63){\makebox(0,0)[lb]{\raisebox{0pt}[0pt][0pt]{\shortstack[l]{{\twlrm OUT}}}}}
\put(67,86){\makebox(0,0)[lb]{\raisebox{0pt}[0pt][0pt]{\shortstack[l]{{\twlrm 2}}}}}
\put(67,46){\makebox(0,0)[lb]{\raisebox{0pt}[0pt][0pt]{\shortstack[l]{{\twlrm 2}}}}}
\put(67,6){\makebox(0,0)[lb]{\raisebox{0pt}[0pt][0pt]{\shortstack[l]{{\twlrm 3}}}}}
\put(128,46){\makebox(0,0)[lb]{\raisebox{0pt}[0pt][0pt]{\shortstack[l]{{\twlrm 3}}}}}
\put(127,6){\makebox(0,0)[lb]{\raisebox{0pt}[0pt][0pt]{\shortstack[l]{{\twlrm 2}}}}}
\put(187,46){\makebox(0,0)[lb]{\raisebox{0pt}[0pt][0pt]{\shortstack[l]{{\twlrm 3}}}}}
\put(4,64){\makebox(0,0)[lb]{\raisebox{0pt}[0pt][0pt]{\shortstack[l]{{\twlrm IN}}}}}
\end{picture}
$$
\vspace*{-15pt}
\caption[]{\label{f3}HALF-PROPAGATOR.}
\end{figure}
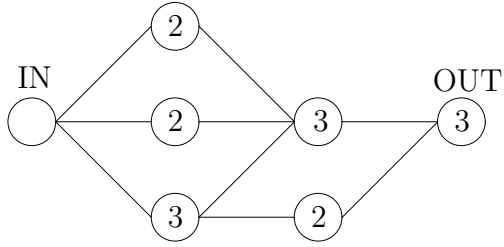

\begin{figure}[t]
%\centerline{\input{fig4.eepic}}
$$
\setlength{\unitlength}{0.0125in}
\begin{picture}(471,163)(0,-10)
\put(300,125){\ellipse{20}{20}}
\put(330,85){\ellipse{20}{20}}
\put(140,45){\ellipse{20}{20}}
\put(110,85){\ellipse{20}{20}}
\put(10,85){\ellipse{20}{20}}
\put(460,125){\ellipse{20}{20}}
\put(430,85){\ellipse{20}{20}}
\put(300,45){\ellipse{20}{20}}
\put(270,85){\ellipse{20}{20}}
\put(170,85){\ellipse{20}{20}}
\put(140,125){\ellipse{20}{20}}
\put(380,85){\ellipse{80}{30}}
\put(60,85){\ellipse{80}{30}}
\put(220,85){\ellipse{80}{30}}
\drawline(340,75)(340,75)
\path(290,125)(270,95)
\path(270,75)(290,45)
\path(330,75)(310,45)
\path(450,125)(430,95)
\path(170,75)(150,45)
\path(110,75)(130,45)
\path(130,125)(110,95)
\path(380,5)(255,5)
\path(60,5)(185,5)
\blacken\path(60,10)(50,5)(60,0)(60,10)
\path(60,10)(50,5)(60,0)(60,10)
\blacken\path(380,10)(390,5)(380,0)(380,10)
\path(380,10)(390,5)(380,0)(380,10)
\put(457,120){\makebox(0,0)[lb]{\raisebox{0pt}[0pt][0pt]{\shortstack[l]{{\twlrm 2}}}}}
\put(427,81){\makebox(0,0)[lb]{\raisebox{0pt}[0pt][0pt]{\shortstack[l]{{\twlrm 3}}}}}
\put(327,81){\makebox(0,0)[lb]{\raisebox{0pt}[0pt][0pt]{\shortstack[l]{{\twlrm 2}}}}}
\put(297,121){\makebox(0,0)[lb]{\raisebox{0pt}[0pt][0pt]{\shortstack[l]{{\twlrm 2}}}}}
\put(297,41){\makebox(0,0)[lb]{\raisebox{0pt}[0pt][0pt]{\shortstack[l]{{\twlrm 2}}}}}
\put(267,81){\makebox(0,0)[lb]{\raisebox{0pt}[0pt][0pt]{\shortstack[l]{{\twlrm 3}}}}}
\put(167,81){\makebox(0,0)[lb]{\raisebox{0pt}[0pt][0pt]{\shortstack[l]{{\twlrm 2}}}}}
\put(137,120){\makebox(0,0)[lb]{\raisebox{0pt}[0pt][0pt]{\shortstack[l]{{\twlrm 2}}}}}
\put(137,41){\makebox(0,0)[lb]{\raisebox{0pt}[0pt][0pt]{\shortstack[l]{{\twlrm 2}}}}}
\put(107,80){\makebox(0,0)[lb]{\raisebox{0pt}[0pt][0pt]{\shortstack[l]{{\twlrm 3}}}}}
\put(4,99){\makebox(0,0)[lb]{\raisebox{0pt}[0pt][0pt]{\shortstack[l]{{\twlrm IN}}}}}
\put(127,139){\makebox(0,0)[lb]{\raisebox{0pt}[0pt][0pt]{\shortstack[l]{{\twlrm OUT}}}}}
\put(287,139){\makebox(0,0)[lb]{\raisebox{0pt}[0pt][0pt]{\shortstack[l]{{\twlrm OUT}}}}}
\put(447,139){\makebox(0,0)[lb]{\raisebox{0pt}[0pt][0pt]{\shortstack[l]{{\twlrm OUT}}}}}
\put(30,82){\makebox(0,0)[lb]{\raisebox{0pt}[0pt][0pt]{\shortstack[l]{{\twlrm propagator}}}}}
\put(189,81){\makebox(0,0)[lb]{\raisebox{0pt}[0pt][0pt]{\shortstack[l]{{\twlrm propagator}}}}}
\put(350,82){\makebox(0,0)[lb]{\raisebox{0pt}[0pt][0pt]{\shortstack[l]{{\twlrm propagator}}}}}
\put(191,1){\makebox(0,0)[lb]{\raisebox{0pt}[0pt][0pt]{\shortstack[l]{{\twlrm any length}}}}}
\end{picture}
$$
\vspace*{-15pt}
\caption[]{\label{f4}MULTIOUTPUT PROPAGATOR.}
\end{figure}
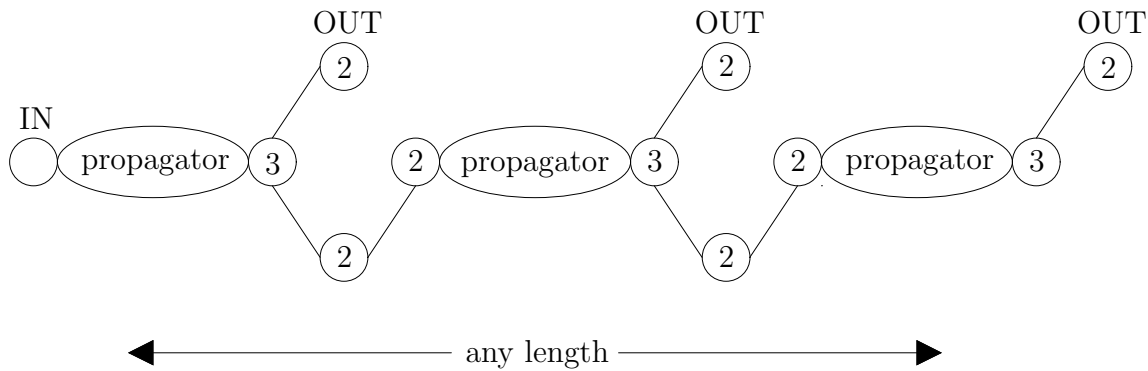

The basic ideas of constructs for the graph involve
"propagators", "half-propagators", "multioutput propagators", and
"initial graphs", with some nodes designated as input nodes,
and some nodes designated as output nodes.
In the following figures a number on a node will be the value $f$ takes
on that node when $G$ is formed.
The value on an \underline{in} node will be acquired when it gets merged
with an \underline{out} node.
A half-propagator is the graph is Fig.~\ref{f3}.
A propagator can be made by merging the \underline{out} node of any
half-propagator with the \underline{in} node of any other half-propagator.
A multioutput propagator is shown in Fig.~\ref{f4}. 
The initial graphs are the graphs in Fig.~\ref{f5} and~\ref{f6}.
\begin{figure}[t]
%\centerline{\input{fig5.eepic}}
$$
\setlength{\unitlength}{0.0125in}
\begin{picture}(144,47)(0,-10)
\put(12,10){\ellipse{20}{20}}
\put(132,10){\ellipse{20}{20}}
\put(72,10){\ellipse{20}{20}}
\path(22,10)(62,10)
\path(82,10)(122,10)
\put(0,23){\makebox(0,0)[lb]{\raisebox{0pt}[0pt][0pt]{\shortstack[l]{{\twlrm OUT}}}}}
\put(69,6){\makebox(0,0)[lb]{\raisebox{0pt}[0pt][0pt]{\shortstack[l]{{\twlrm 2}}}}}
\put(9,6){\makebox(0,0)[lb]{\raisebox{0pt}[0pt][0pt]{\shortstack[l]{{\twlrm 2}}}}}
\put(130,6){\makebox(0,0)[lb]{\raisebox{0pt}[0pt][0pt]{\shortstack[l]{{\twlrm 2}}}}}
\put(120,23){\makebox(0,0)[lb]{\raisebox{0pt}[0pt][0pt]{\shortstack[l]{{\twlrm OUT}}}}}
\end{picture}
$$
\vspace*{-15pt}
\caption[]{\label{f5}A "$\exists$-graph".}
\end{figure}
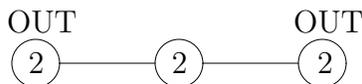
\begin{figure}[t]
%\centerline{\input{fig6.eepic}}
$$
\setlength{\unitlength}{0.0125in}
\begin{picture}(263,115)(0,-10)
\put(72,50){\ellipse{20}{20}}
\put(192,50){\ellipse{20}{20}}
\put(12,50){\ellipse{20}{20}}
\put(252,50){\ellipse{20}{20}}
\put(132,10){\ellipse{20}{20}}
\put(132,90){\ellipse{20}{20}}
\path(22,50)(62,50)
\path(122,90)(72,60)
\path(72,40)(122,10)
\path(192,40)(142,10)
\path(142,90)(192,60)
\path(242,50)(202,50)
\put(239,64){\makebox(0,0)[lb]{\raisebox{0pt}[0pt][0pt]{\shortstack[l]{{\twlrm OUT}}}}}
\put(190,46){\makebox(0,0)[lb]{\raisebox{0pt}[0pt][0pt]{\shortstack[l]{{\twlrm 2}}}}}
\put(249,47){\makebox(0,0)[lb]{\raisebox{0pt}[0pt][0pt]{\shortstack[l]{{\twlrm 2}}}}}
\put(130,6){\makebox(0,0)[lb]{\raisebox{0pt}[0pt][0pt]{\shortstack[l]{{\twlrm 2}}}}}
\put(129,86){\makebox(0,0)[lb]{\raisebox{0pt}[0pt][0pt]{\shortstack[l]{{\twlrm 2}}}}}
\put(70,46){\makebox(0,0)[lb]{\raisebox{0pt}[0pt][0pt]{\shortstack[l]{{\twlrm 2}}}}}
\put(0,64){\makebox(0,0)[lb]{\raisebox{0pt}[0pt][0pt]{\shortstack[l]{{\twlrm OUT}}}}}
\put(9,46){\makebox(0,0)[lb]{\raisebox{0pt}[0pt][0pt]{\shortstack[l]{{\twlrm 2}}}}}
\end{picture}
$$
\vspace*{-15pt}
\caption[]{\label{f6}A "$\forall$-graph".}
\end{figure}

The graph $G$ consists of the following.
For each $i$ from $1$ to $k$, we have a $\forall$-graph, with the 
\underline{out} nodes named $U_i$ and $\overline{U}_i$.
For each $i$ from $k+1$ to $k+r$, we have a $\exists$-graph, with the
\underline{out} nodes names $U_i$ and $\overline{U}_i$.
We think of the $C_i$'s as clauses, 
and think of $U_s$ and $\overline{U}_s$ as literals.
For each literal $V$ we connect a multioutput propagator to the node
named $V$, identifying the \underline{in} node of the propagator
with $V$.
All the multioutput propagators look alike having $3m$ output nodes,
one for each $ij$ where $1 \leq i \leq m$ and $1 \leq j \leq 3$.

Now we add $m$ new nodes (each with $f(C_i)=3$) named 
$C_1,C_2, \ldots ,C_m$.
For each $i$ from $1$ to $m$, and each $j$ from $1$ to $3$,     
connect $C_i$ to the $ij$ node of the multioutput propagator
attached to the node named~$X_{ij}$.

That describes the graph $G$, which is obviously bipartite.
Every variable occurs in at most three clauses, and therefore
it occurs either at most once positive or
at most once negative.
Combining this with the fact that $G_{\Phi}$ is planar, we
conclude that $G$ is planar.

We use here a different half-propagator from the one used 
in~\cite{ERT}, and therefore the following properties needed for 
the proof should be verified for our half-propagator.
\begin{enumerate}
\item
A $2$-coloration will give the \underline{out} node opposite color
to that of the \underline{in} node.
\item
For any choice of a letter from the \underline{in} node, and no
matter what letters are put on nodes other than the \underline{in}
node, there is a compatible choice of letters from the remaining 
nodes of the half-propagator.
\item
For any assignment of letters to nodes other than the \underline{in}
node, for any choice of a letter from the \underline{out} node, there
is at most one choice of letter incompatible with it on the
\underline{in} node. 
(This is a direct consequence of $K_{2,3}$ being $2$-choosable)
\item
\label{s4}
There is an assignment of letters, and a choice of \underline{in}
letter, such that only one choice of a letter from the \underline{out}
node is compatible with it. (See Fig.~\ref{f7})
\end{enumerate}
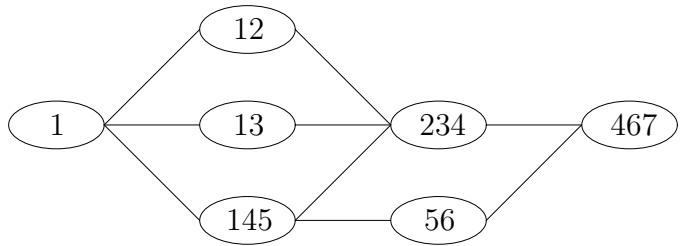
\begin{figure}[t]
%\centerline{\input{fig7.eepic}}
$$
\setlength{\unitlength}{0.0125in}
\begin{picture}(280,115)(0,-10)
\put(100,50){\ellipse{40}{20}}
\put(180,50){\ellipse{40}{20}}
\put(100,90){\ellipse{40}{20}}
\put(100,10){\ellipse{40}{20}}
\put(260,50){\ellipse{40}{20}}
\put(20,50){\ellipse{40}{20}}
\put(180,10){\ellipse{40}{20}}
\path(80,90)(40,50)
\path(40,50)(80,50)
\path(40,50)(80,10)
\path(120,90)(160,50)
\path(160,50)(120,50)
\path(160,50)(120,10)
\path(240,50)(200,50)
\path(240,50)(200,10)
\path(120,10)(160,10)
\put(172,46){\makebox(0,0)[lb]{\raisebox{0pt}[0pt][0pt]{\shortstack[l]{{\twlrm 234}}}}}
\put(17,46){\makebox(0,0)[lb]{\raisebox{0pt}[0pt][0pt]{\shortstack[l]{{\twlrm 1}}}}}
\put(94,86){\makebox(0,0)[lb]{\raisebox{0pt}[0pt][0pt]{\shortstack[l]{{\twlrm 12}}}}}
\put(94,46){\makebox(0,0)[lb]{\raisebox{0pt}[0pt][0pt]{\shortstack[l]{{\twlrm 13}}}}}
\put(91,6){\makebox(0,0)[lb]{\raisebox{0pt}[0pt][0pt]{\shortstack[l]{{\twlrm 145}}}}}
\put(174,6){\makebox(0,0)[lb]{\raisebox{0pt}[0pt][0pt]{\shortstack[l]{{\twlrm 56}}}}}
\put(252,46){\makebox(0,0)[lb]{\raisebox{0pt}[0pt][0pt]{\shortstack[l]{{\twlrm 467}}}}}
\end{picture}
$$
\vspace*{-15pt}
\caption[]{\label{f7}An assignment for the half-propagator.}
\end{figure}

The proof which appears in~\cite{ERT} can be used to
conclude that $G$ is $f$-choosable iff $B$ is true. $\Box$

\section{The choosability of planar graphs}

In this section we prove Theorems~\ref{t110} and~\ref{t111}.              
\begin{lemma}
\label{l41}
Let $G=(V,E)$ be an odd cycle, 
and suppose we have an assignment of sets of integers 
$S(v) \subseteq Z$
for all vertices $v \in V$, where $S(v)=2$ for all $v$.  
There exists a proper coloring $c:V \mapsto Z$ so that $c(v) \in S(v)$
for all $v \in V$ if and only if not all the sets $S(v)$ are equal.
\end{lemma}
{\bf Proof}\,
Suppose first that not all the sets $S(v)$ are equal.
Let $x_1$ and $x_k$ be adjacent vertices for which $S(x_1) \neq S(x_k)$,
where $G$ is the cycle $x_1-\cdots-x_k-x_1$.
Choose a color $c_1 \in S(x_1)-S(x_k)$, and go in a sequence choosing
$c_2 \in S(x_2)-\{c_1\}$, $c_3 \in S(x_3)-\{c_2\},\ldots$ until
$c_k \in S(x_k)-\{c_{k-1}\}$. We have obtained a proper coloring of
$G$, as needed.

If the sets $S(v)$ are equal
there is no coloring as $\chi(G)=3$. $\Box$
\begin{lemma}
\label{l42}
Suppose that $C_1$ and $C_2$ are two disjoint copies of the 
odd cycle of length $k$, which we denote by $C_1=x_1-\cdots-x_k-x_1$ and
$C_2=y_1-\cdots-y_k-y_1$.
Let $G$ be composed of $C_1$ and $C_2$ together with the edges 
$(x_i,y_i)$, $i=1,\ldots,k$.
Suppose we have an assignment of sets of integers 
$S(v) \subseteq Z$ for all vertices $v \in C_2$, 
where $S(v)=3$ for all $v \in C_2$.
Then there is at most one proper coloring of $C_1$
which cannot be completed to a proper coloring of $G$ by 
assigning to each vertex $v \in C_2$ a color from $S(v)$.  
\end{lemma}
{\bf Proof}\,
Suppose that $c$ is a proper coloring of $C_1$ which cannot be completed
to a proper coloring of $G$. Denote $c_i=c(x_i)$, $i=1,\ldots,k$.
If follows from lemma~\ref{l41} that there exist two colors $a$
and $b$ so that $S(y_i)=\{a,b,c_i\}$, $i=1,\ldots,k$.
Since $c$ is a proper coloring, surely $\cap_{i=1}^{k}S(y_i)=\{a,b\}$.
By applying lemma~\ref{l41} again, we conclude 
that $c$ is the only proper coloring with the required properties
for the considered assignment of sets $S(y_i)=\{a,b,c_i\}$. $\Box$
\begin{defi}
A graph $G=(V,E)$ is {\em $k$-restrictly-choosable} if $G$ is $f_v$-choosable
for every $v \in V$, where the function $f_v$ is defined as $f_v(v)=k-1$
and $f_v(w)=k$ for every $w \in V-\{v\}$.
\end{defi}
\begin{defi}
A graph $G$ is {\em $k$-choice-critical} if $G$ is $k$-choosable
but not $k$-restrictly-choosable.
\end{defi} 
\begin{defi}
Let $G=(V,E)$ be a graph, and suppose that $u$ and $v$ are two distinct
vertices of $G$.               
Let $S$ be an assignment of sets of integers $S(w) \subseteq Z$
for all vertices $w \in V$.   
We denote by $incomp(G,u,v,S)$ the set
$\{(a,b) \in S(u) \times S(v)|$ 
there is no proper vertex coloring 
$c:V \mapsto Z$ so that $c(u)=a$, $c(v)=b$ and 
$c(w) \in S(w)$ for all $w \in V\}$.
\end{defi}
\begin{lemma}
\label{l43}
Let $W_2=(V,E)$ be the graph in Fig.~\ref{f2}.   
If $S$ is an assignment of sets of integers $S(w) \subseteq Z$ for 
all vertices $w \in V$, where $S(w)=3$ for all $w$,     
then $|incomp(W_2,u,v,S)| \leq 1$. 
\end{lemma}
{\bf Proof}\,
Suppose that $(a,b) \in incomp(W_2,u,v,S)$. It is easy to verify,
by applying lemma~\ref{l42}, that $a \neq b$, 
$a \in S(w)$, $b \in S(x_4) \cap S(y_4)$ and 
$\{a,b\} \subseteq S(x_2) \cap S(y_2)$.
Combining lemma~\ref{l42} with the fact that 
$(a,b) \in incomp(W_2,u,v,S)$, we obtain that there exist 
a coloring of the vertices $x_1,\ldots,x_4,w$ with the colors
$c_1,\ldots,c_5$, respectively, and a coloring of the vertices
$y_1,\ldots,y_4,w$ with the colors $d_1,\ldots,d_5$, respectively,
which have the properties stated in the lemma.
It follows easily that $S(w)=\{a,c_5,d_5\}$ and 
$S(x_2)=\{a,b,c_2\}$.

In the same manner we can prove that if $(g,h) \in incomp(W_2,u,v,S)$,
then $g \neq h$, $S(w)=\{g,c_5,d_5\}$ and $S(x_2)=\{g,h,c_2\}$, which
implies that $g=a$ and $h=b$.
This proves that $|incomp(W_2,u,v,S)| \leq 1$, as needed. $\Box$

We construct the graph $H_1$ as follows: 
We take the disjoint union of the graphs $\{G_i:1 \leq i \leq 6\}$,
where each $G_i$ is a copy of the graph $W_2$ in Fig.~\ref{f2}.
All the $6$ vertices named $u$ are identified, as well as all the 
$6$ vertices named $v$, to obtain the planar triangle-free graph $H_1$.
\begin{lemma}
\label{l44}
The graph $H_1$ is $3$-choosable.
\end{lemma}
{\bf Proof}\,
Let $S$ be an assignment of sets of integers $S(w) \subseteq Z$ for
all vertices $w \in V$, where $S(w)=3$ for all~$w$.
Suppose first that there exists a color $c \in S(u) \cap S(v)$.
It follows immediately that by coloring $u$ and $v$ with 
the color $c$ we can find a proper coloring.

Suppose next that $S(u) \cap S(v)=\emptyset$.
It follows from lemma~\ref{l43} that $|incomp(G_i,u,v,S)| \leq 1$   
for $i=1,\ldots,6$, and therefore $|incomp(H_1,u,v,S)| \leq 6$.
Since $|incomp(H_1,u,v,S)| < |S(u) \times S(v)|=9$, we conclude
that a coloring in possible. $\Box$
\begin{lemma}
\label{l45}
The graph $H_1$ is not $3$-restrictly-choosable.
\end{lemma}
{\bf Proof}\,
Take $S(u)=\{10,11\}$ and $S(v)=\{12,13,14\}$. 
Proceed as in the proof of Theorem~\ref{t18}. $\Box$
\begin{lemma}
\label{l46}
There exists a planar triangle-free graph which is $3$-choice-critical.
\end{lemma}
{\bf Proof}\,
Combine lemmas~\ref{l44} and~\ref{l45}. $\Box$ 

\noindent
{\bf Proof of Theorem~\ref{t110}}\,
It is easy to see that {\bf PTFG $3$-CH}$\in \Pi_2^p$.
We transform {\bf BPG $(2,3)$-CH} to {\bf PTFG $3$-CH}.
Let the graph $G=(V,E)$ and the function $f:V \mapsto \{2,3\}$
be an instance of {\bf BPG $(2,3)$-CH}.
We shall construct a planar triangle-free graph $G'=(V',E')$ such that
$G'$ is $3$-choosable if and only if $G$ is $f$-choosable.
If follows from lemma~\ref{l46} that there exists a planar triangle-free
graph $W$ which is $3$-choice-critical.
Let $u$ be a vertex of $W$ for which $W$ is not $g_u$-choosable, where
the function $g$ is defined as $g_u(u)=2$ and $g_u(w)=3$ otherwise. 
The graph $G'$ is obtained from $G$ by adding a disjoint copy of $W$
for every $v \in V(G)$ for which $f(v)=2$, and connecting $v$ to the
vertex $u$ of this copy.

Since both $G$ and $W$ are planar triangle-free graphs, it is easy to
see that $G'$ is also a planar triangle-free graph
(recall that $W$ has an embedding in the plane so that $u$ appears
on the exterior face.) 
We first prove that if $G$ is $f$-choosable, then $G'$ is $3$-choosable.
Take an assignment of sets of integers $S(w) \subseteq Z$ for all
vertices $w \in V'$, where $S(w)=3$ for all $w$.
The graph $W$ is $3$-choosable, and so we find a proper coloring in
each copy of $W$ in the graph $G'$. For each copy of $W$, the 
color chosen in the vertex $u$ is removed from the vertex of $G$ 
adjacent to $u$. The coloring can be completed, since $G$ is
$f$-choosable.

We now prove that if $G'$ is $3$-choosable, then $G$ is $f$-choosable.
Suppose we have an assignment of sets of integers $S(w) \subseteq Z$ for all
vertices $w \in V(G)$, where $|S(w)|=f(w)$ for all $w$.
Take an assignment which proves that $W$ is not $g_u$-choosable,
and put it in each copy of $W$ in the graph $G'$.
Let $d$ be a new color. For each copy $W$, we add the color $d$
to the vertex $u$ of this copy and to its neighbor in $G$.
Since $G'$ is $3$-choosable, we can find a proper coloring $c$ of $G'$
assigning to each vertex a color from its set.
The coloring $c$ restricted to $G$ implies that $G$ is $f$-choosable. 
$\Box$

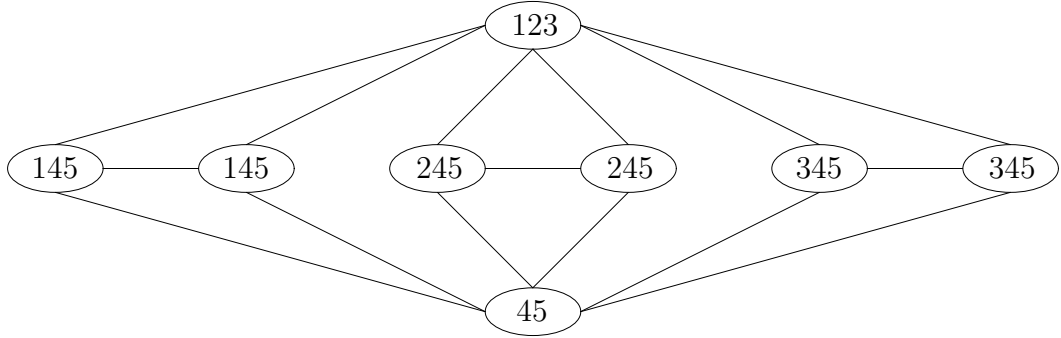
\begin{figure}[t]
%\centerline{\input{fig8.eepic}}
$$
\setlength{\unitlength}{0.0125in}
\begin{picture}(440,155)(0,-10)
\put(260,70){\ellipse{40}{20}}
\put(180,70){\ellipse{40}{20}}
\put(100,70){\ellipse{40}{20}}
\put(20,70){\ellipse{40}{20}}
\put(220,130){\ellipse{40}{20}}
\put(220,10){\ellipse{40}{20}}
\put(340,70){\ellipse{40}{20}}
\put(420,70){\ellipse{40}{20}}
\path(200,130)(100,80)
\path(200,130)(20,80)
\path(20,60)(200,10)
\path(100,60)(200,10)
\path(80,70)(40,70)
\path(220,120)(180,80)
\path(180,60)(220,20)
\drawline(220,120)(220,120)
\path(220,120)(260,80)
\path(260,60)(220,20)
\path(200,70)(240,70)
\path(360,70)(400,70)
\path(240,130)(340,80)
\path(240,130)(420,80)
\path(340,60)(240,10)
\path(420,60)(240,10)
\put(213,6){\makebox(0,0)[lb]{\raisebox{0pt}[0pt][0pt]{\shortstack[l]{{\twlrm 45}}}}}
\put(251,66){\makebox(0,0)[lb]{\raisebox{0pt}[0pt][0pt]{\shortstack[l]{{\twlrm 245}}}}}
\put(330,66){\makebox(0,0)[lb]{\raisebox{0pt}[0pt][0pt]{\shortstack[l]{{\twlrm 345}}}}}
\put(411,66){\makebox(0,0)[lb]{\raisebox{0pt}[0pt][0pt]{\shortstack[l]{{\twlrm 345}}}}}
\put(211,126){\makebox(0,0)[lb]{\raisebox{0pt}[0pt][0pt]{\shortstack[l]{{\twlrm 123}}}}}
\put(171,66){\makebox(0,0)[lb]{\raisebox{0pt}[0pt][0pt]{\shortstack[l]{{\twlrm 245}}}}}
\put(90,66){\makebox(0,0)[lb]{\raisebox{0pt}[0pt][0pt]{\shortstack[l]{{\twlrm 145}}}}}
\put(10,66){\makebox(0,0)[lb]{\raisebox{0pt}[0pt][0pt]{\shortstack[l]{{\twlrm 145}}}}}
\end{picture}
$$
\vspace*{-15pt}
\caption[]{\label{f8}The graph $W_3$.}
\end{figure}

In order to prove that deciding whether a given planar graph
is $3$-choosable is $\Pi_2^p$-complete 
(a weaker version of Theorem~\ref{t110}),
it is possible to use the planar graph $W_3$ in Fig.~\ref{f8}. 
In a similar manner to the previous proofs, one can prove that
$W_3$ is $3$-choice-critical.
The assignment given in Fig.~\ref{f8} proves that $W_3$ is not
$3$-restrictly-choosable.

\begin{lemma}
\label{l47}
Let $W_1=(V,E)$ be the graph in Fig.~\ref{f1}.   
If $S$ is an assignment of sets of integers $S(w) \subseteq Z$ for 
all vertices $w \in V$, where $S(w)=4$ for all $w$,     
then $|incomp(W_1,u,v,S)| \leq 1$. 
\end{lemma}
{\bf Proof}\,
Suppose that $\{a,b\}\in incomp(W_1,u,v,S)$.
It is easy to verify, by applying lemma~\ref{l41}, that
$a \neq b$, $a \in S(x_1) \cap S(y_1)$, $b \in S(x_3) \cap S(y_3)$
and $\{a,b\} \subseteq S(w) \cap S(x_2) \cap S(y_2)$.
Combining lemma~\ref{l41} with the fact that $\{a,b\}\in incomp(W_1,u,v,S)$,
we obtain that there exist three distinct colors $c$,$d$ and $e$ so
that $S(x_2)=\{a,b,c,d\}$, $S(x_1)=\{a,c,d,e\}$ and 
$S(x_3)=\{b,c,d,e\}$.

In the same manner we can prove that if $(g,h) \in incomp(W_1,u,v,S)$,
then $g \neq h$, $S(x_1)=\{g,c,d,e\}$ and $S(x_3)=\{h,c,d,e\}$,
which implies that $g=a$ and $h=b$. 
This proves that $|incomp(W_1,u,v,S)| \leq 1$, as needed. $\Box$
\begin{lemma}
\label{l48}
There exists a planar graph which is $4$-choice-critical.
\end{lemma}
{\bf Proof}\,
Take $12$ pairwise disjoint copies of the graph $W_1$ in Fig.~\ref{f1}
and identify all the $12$ vertices named $u$ as well as
all the $12$ vertices named $v$.
Use lemma~\ref{l47} and proceed as in the proofs 
of lemmas~\ref{l44} and~\ref{l45}.
$\Box$

\noindent
{\bf Proof of Theorem~\ref{t111}}\,
Apply lemma~\ref{l48} as in the proof of Theorem~\ref{t110}. $\Box$

\noindent
{\bf Acknowledgement}\,
I would like to thank Noga Alon and Michael Tarsi for helpful discussions.

\end{document}